\documentclass[runningheads]{llncs}
\usepackage[colorlinks,linkcolor=blue]{hyperref}
\usepackage{graphicx}
\usepackage{cite}
\usepackage{amsmath,amssymb,amsfonts}
\usepackage{bm}
\usepackage{algpseudocode}
\usepackage{graphicx}
\usepackage{textcomp}
\usepackage{xcolor}
\usepackage{multirow}
\usepackage{booktabs}
\usepackage{caption2}
\usepackage{subfigure} %子图
\usepackage{threeparttable} 

\usepackage{wrapfig}
\usepackage[ruled,vlined]{algorithm2e}
\newcommand{\corr}{(\Letter)}

\usepackage[T1]{fontenc}
\usepackage[misc]{ifsym}
\usepackage{mwe}
\begin{document}
\title{Towards Open-World Cross-Domain Sequential Recommendation: A Model-Agnostic Contrastive Denoising Approach}
\titlerunning{Open-World CDSR: A Model-Agnostic Contrastive Denoising Approach}
% \author{Submission ID: 88}
%
% \authorrunning{Anonymous et al.}
% If the full title of your paper is short enough to also fit in the running head, you can omit the abbreviated paper title here. You can check as follows: if you comment out the \titlerunning line, something will appear in the header of all odd-numbered pages of your PDF from page 3 onward. This something is either the full title (in which case all is well), or the error message "Title Suppressed Due to Excessive Length". If this error message appears, you're going to want to provide an abbreviated title within the \titlerunning command, because if you won't do it, Springer will do it for you.

%N.B.: Author information (both in the \author{} and \authorrunning{} command) should only be present in the Camera-Ready Version of your paper. The version that you initially submit for review, ought to be double-blind. So, when initially submitting your paper, use:
%\author{Author information scrubbed for double-blind reviewing}
\author{Wujiang Xu\inst{1} \and
Xuying Ning\inst{2} \and
Wenfang Lin\inst{4} \and 
Mingming Ha \inst{4} \and
Qiongxu Ma \inst{4} \and
Qianqiao Liang \inst{4} \and 
Xuewen Tao \inst{4} \and
Linxun Chen \inst{4} \and 
Bing Han \inst{4} \and \\
Minnan Luo  \corr \inst{3} }

% You may leave out the orcidID information, if you want to.
% Use \corr to indicate the corresponding author. Note the spacing around the \corr command. Only one author can be the corresponding author.

%N.B.: comment out the \authorrunning{} command for the double-blind version of your paper submitted for review. Later, if your paper is accepted, use the command for the Camera-Ready Version.
\authorrunning{Xu et al.}
% If there are two authors, write 'A.L. Benjamin and C.C. Broadus Jr.'
% If there are more than two authors, '[...] et al.' is used.

\institute{Department of Computer Science, Rutgers University, NJ 08854, US
\and
Department of Computer Science, University of Illinois
Urbana-Champaign, US
\and
School of Computer Science and Technology, Xi’an Jiaotong University, Xi’an 710049, China \email{minnluo@xjtu.edu.cn}
\and
MYbank, Ant Group, Hangzhou, China}

\maketitle              % typeset the header of the contribution
\vspace{-20pt}

\begin{abstract}
Cross-domain sequential recommendation (CDSR) aims to address the data sparsity problems that exist in traditional sequential recommendation (SR) systems. 
  Existing CDSR approaches aim to design a specific cross-domain unit to transfer information by relying on overlapping users. 
  However, in real-world recommender systems, CDSR scenarios typically involve a majority of long-tailed users exhibiting sparse behaviors and cold-start users who are solely present within a single domain. 
  Consequently, lacking exploration of these users' interests, existing CDSR methods perform poorly on real-world platforms. Therefore, it becomes imperative to explore complementary interest information to enhance the model's performance  (\textit{1st} CH).
  Recently, some SR approaches have utilized auxiliary behaviors to complement the information for long-tailed users. However, these methods cannot deliver promising performance in CDSR, as they overlook the semantic gap between target and auxiliary behaviors, as well as user interest deviation across domains (\textit{2nd} CH). In this paper, we propose a model-agnostic contrastive denoising (MACD) approach towards open-world CDSR. 
  We introduce auxiliary behavior sequence information (i.e., clicks) into CDSR methods to explore potential interests. 
  Specifically, we design a denoising interest-aware network combined with a contrastive information regularizer to remove inherent noise from auxiliary behaviors and exploit multi-interest from users. Extensive offline experiments on public industry datasets and a standard A/B test on a large-scale financial platform with millions of users both confirm the remarkable performance of our model in open-world CDSR scenarios. Code and dataset are available at \href{https://github.com/WujiangXu/MACD}{URL}.

\end{abstract}
\section{Introduction}
Sequential Recommendation (SR) \cite{hidasi2015session,kang2018self,sun2019bert4rec,tang2018personalized} has gained significant attention in recent years due to its ability to model dynamic user preferences.
However, in real-world platforms, users often exhibit partial behaviors within specific domains, leading to biased observed preferences based on single-domain interactions. To address this issue, cross-domain sequential recommendation (CDSR) methods \cite{ma2019pi,ma2022mixed,cao2022contrastive,li2021dual} have been proposed to construct models using multi-domain data, transferring rich information from other relevant domains to improve performance in multiple domains simultaneously. 
To transfer the information across domains, the core idea of CDSR works is to extend the single-domain sequential recommendation methods via designing a mapping function. Specifically, PiNet \cite{ma2019pi} employs a designed gating mechanism to filter information and propagate this information for overlapping users from the source domain to the target domain. Similarly, DASL \cite{li2021dual} proposes a dual-attention mechanism to extract the user-shared information for each domain. To enhance the correlation between single- and cross-domain user preferences, C$^2$DSR \cite{cao2022contrastive} combines an attention information transferring unit with a contrastive infomax objective.

% \begin{figure}[tb!]
% \centering{
% \includegraphics[width=0.45\textwidth]{Figure/intro.pdf}}
% % \captionsetup{font={small}}
% \caption{Previous methods focus on constructing their structure based on overlapping users with rich behaviours (a) under a closed-world environment. In this study, we aim to design the model for an open-world environment that accounts for the majority of long-tailed users (b) and cold-start users (c) with sparse historical behaviours. }
% \label{intro}
% \end{figure}

\begin{figure}[tb!]
    \begin{minipage}{0.55\textwidth}
        \centering
        \includegraphics[width=\textwidth]{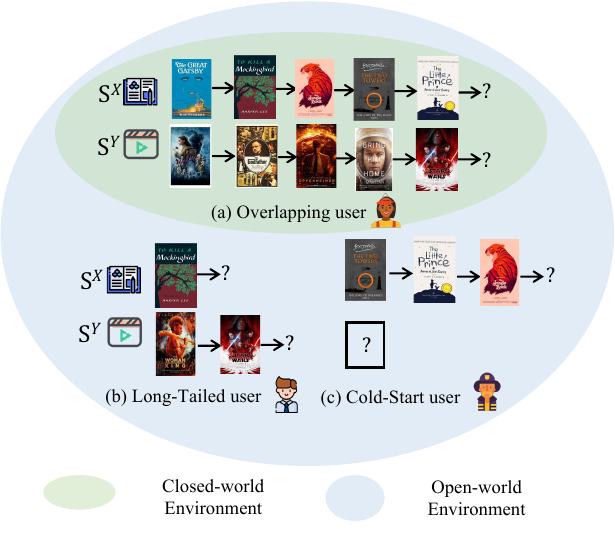}
        \label{intro}
    \end{minipage}
    \hfill % optional: to add some spacing between the image and the caption
    \begin{minipage}{0.43\textwidth}
        \caption{Previous methods focus on constructing their structure based on overlapping users with rich behaviours (a) under a closed-world environment. In this study, we aim to design the model for an open-world environment that accounts for the majority of long-tailed users (b) and cold-start users (c) with sparse historical behaviours.}
    \end{minipage}
\end{figure}

Despite the promising improvements, these mapping function designs of existing CDSR methods heavily rely on overlapping users with rich behaviors, which can lead to unsatisfactory performance in an open-world environment \cite{wu2021towards,wu2021towards2}. Typically, these methods assume full or majority overlap of users (Fig. \ref{intro}(a)) across domains, which is only a minority in real-world platforms such as Taobao or Amazon. These approaches disregard most users in the open-world CDSR environment, particularly the long-tailed users (Fig. \ref{intro}(b)) with few interactions and the cold-start users (Fig. \ref{intro}(c)) who are only present in a single domain. As a result, these approaches fail to provide sufficient exploration of cross-domain information to long-tailed and cold-start users, leading to an incomplete understanding of their interests.
Therefore, the first challenge is: \textit{How can we explore complementary interest information to enhance the model's performance
towards open-world CDSR scenarios, where the majority of users are long-tailed or cold-start users with sparse historical behaviors?}

Recent SR studies \cite{jin2020multi,xia2021graph} have explored the integration of multi-typed behaviors to improve the performance of long-tailed users. In real-world platforms, users interact with items through multiple types of behavior. 
The target behavior (e.g. purchase) that directly benefits businesses is usually sparser than other behavior types (e.g. view or click). To capture behavior semantics, MBGCN \cite{jin2020multi} designs a user-item and item-item graph neural network, while MBHT \cite{yang2021hyper} proposes a hypergraph-based transformer to encode behavior-aware sequential patterns from both fine-grained and coarse-grained levels. More recently, DPT \cite{zhang2023denoising} develops a denoising and prompt-tuning framework to guide the informative and relative representations. However, these multi-behavior denoising SR approaches cannot fully address the CDSR problem as they fail to consider the deviation of user interests across domains and neglect the semantic gap between auxiliary and target behaviors. Therefore, the second challenge is: \textit{How can the semantic gap between target and auxiliary behaviors be reduced and user interest deviation be learned when utilizing auxiliary behavior sequences to enhance information for long-tailed users in CDSR?}

To address the aforementioned challenges, we propose a \textbf{M}odel-\textbf{A}gnostic \textbf{C}ontrastive \textbf{D}enoising approach, namely \textbf{MACD}. Overall, our major contributions can be summarized as follows:

(1) We propose a \textbf{m}odel-\textbf{a}gnostic \textbf{c}ontrastive \textbf{d}enoising framework, namely \textbf{MACD}, towards open-world CDSR that can be integrated with most off-the-shelf SR methods. To the best of our knowledge, we are the first who utilize the auxiliary behaviour information in CDSR models, which incorporate informative potential interests of users, especially for long-tailed users and cold-start users.  

(2) We propose a denoising interest-aware network that incorporates an intra-domain / cross-domain denoising module and a contrastive information regularizer. This network aims to reduce the semantic gap between target and auxiliary behaviors, enabling to better capture of user interest.

(3) We introduce a fusion gate unit to enhance the fusion of representations, and we employ a parameter-free inductive representation generator to generate inductive representations for cold-start users during the inference stage.

% By incorporating auxiliary behaviors into our models, we enhance model consistency and effectiveness, and provide complementary information for users, particularly those who are long-tailed and cold-start.

% (2) The designed denoising interest-aware network explores users' explicit and implicit interests and transfers cross-domain knowledge. Additionally, we propose a contrastive information regularizer and fusion gate unit to normalize interest representation learning and reduce the semantic gap between target and auxiliary behaviors. For the cold-start users, we also propose a parameter-free inductive representation generator during the inference stage to obtain an inductive representation.

(4) We conduct extensive experiments on a large-scale scenario with a minority of overlapping users, representing an open-world environment. Furthermore, a standard A/B test is conducted to validate our performance on a real-world CDSR financial platform with millions of daily traffic logs.

\section{Methodology}
\subsection{Problem Formulation}
In this work, we consider a general CDSR scenario in an open-world environment that consists of partially overlapping users, a majority of long-tailed and cold-start users across two domains, namely domain $X$ and domain $Y$. The data is denoted by $D^X=(\mathcal{U}^X,\mathcal{V}^X,\mathcal{E}^X)$ and $D^Y=(\mathcal{U}^Y,\mathcal{V}^Y,\mathcal{E}^Y)$, where $\mathcal{U}$, $\mathcal{V}$, and $\mathcal{E}$ are the sets of users, items, and interaction edges, respectively. For a given user, we denote their target item interaction sequences in chronological order as $S^X=[v^X_1,v^X_2,\cdots,v^X_{|S^X|}]$ and $S^Y=[v^Y_1,v^Y_2,\cdots,v^Y_{|S^Y|}]$, where $|\cdot|$ denotes the length of the sequence. We further introduce the user's auxiliary behavior sequences, $C^X=[v^X_1,v^X_2,\cdots,v^X_{|C^X|}]$ and $C^Y=[v^Y_1,v^Y_2,\cdots,v^Y_{|C^Y|}]$\footnote{Due to the low cost of engagement, the auxiliary behaviors (e.g., clicks or views) provide richer records than the target user-item interactions (e.g., purchases).} to enrich user behaviors and solve the issue that previous CDSR methods being heavily affected by data sparsity problems. The adjusted objective of CDSR in our setting is to predict the next item for a given user based on both user's  target behavior sequence and auxiliary behavior sequence:
%adjusted part!
 \begin{align} 
&\mathrm{max}\;\;P^X\left(v^X_{|S^X|+1}=v|S^X,S^Y,C^X,C^Y\right),\mathrm{if }\;\; v \in \mathcal{V}^X. \notag\\
%\end{equation}
%\begin{equation} 
&\mathrm{max}\;\;P^Y\left(v^Y_{|S^Y|+1}=v|S^X,S^Y,C^X,C^Y\right),\mathrm{if }\;\; v \in \mathcal{V}^Y.\notag
\end{align}

In the cold-start setting, where we have users only observed in one domain (e.g., $Y$) and thus being cold-start in another domain (e.g., $X$), the objective function can be generalized by adding an additional condition that $S^X = \varnothing, C^X = \varnothing$. 
Moreover, we classify users into two categories based on the length of their target sequence. If users' target sequence length, denoted as $ |S^X| $, is less than $L_{head}^X$, they are categorized as long-tailed users; otherwise, they are classified as head users. We determine $L_{head}^{X}, L_{head}^{Y}$ by calculating the average length of the top 20\% longest sequences in the specific domain. It should be noticed that the definition of long-tailed users is independent across different domains. 

\begin{figure}[tb!]
    \begin{minipage}{0.64\textwidth}

\centering{
\includegraphics[width=0.99\textwidth]{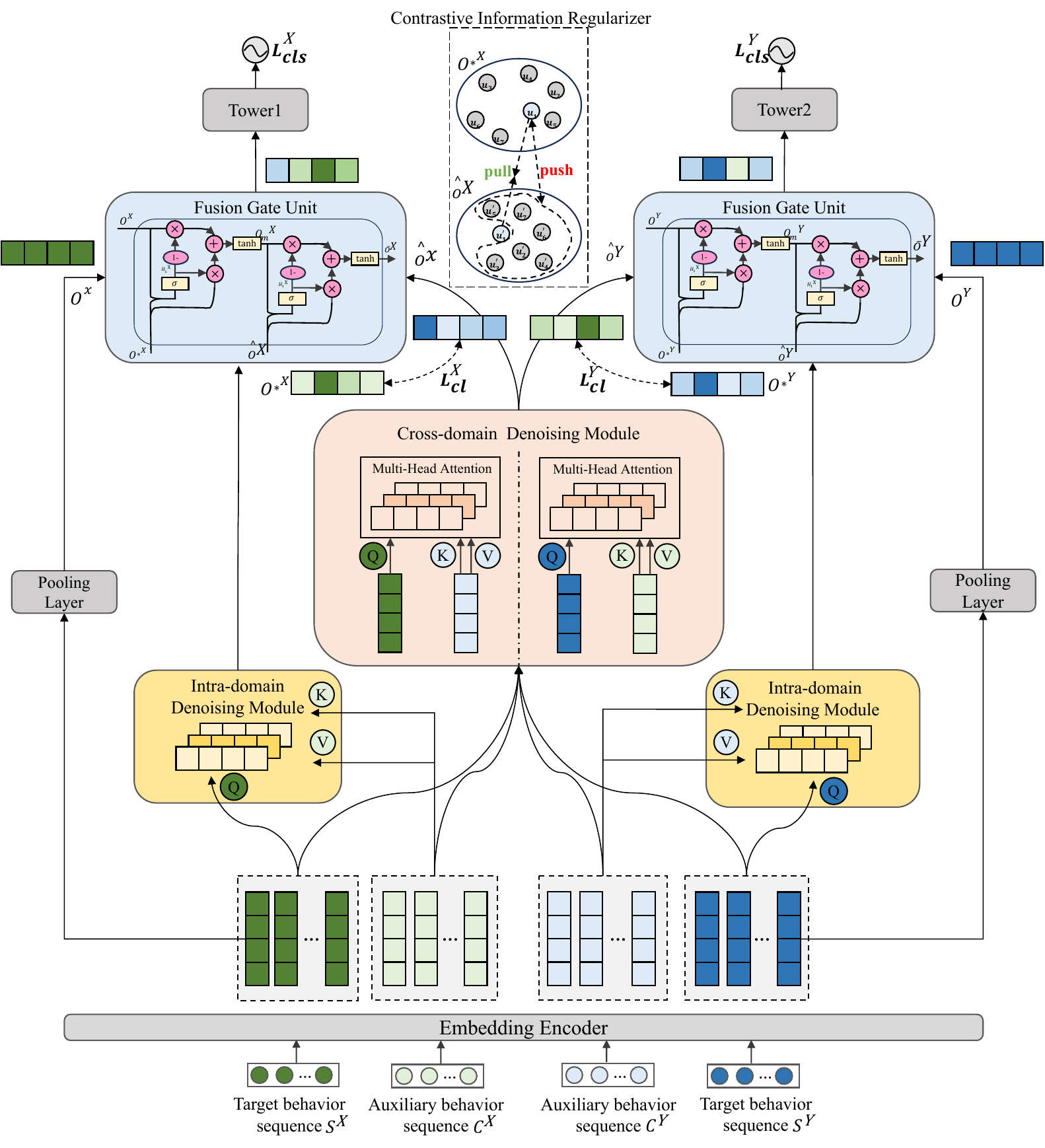}}
% \captionsetup{font={footnotesize}}
\label{Fig_framework}
\end{minipage}
\hfill
\begin{minipage}{0.35\textwidth}
\captionsetup{font={scriptsize}}
    \caption{\textcolor{black}{Overview of our MACD approach.} Unlike previous CDSR methods, our MACD is a general and model-agnostic approach that can be integrated with most off-the-shelf SDSR methods. Our MACD fully leverages auxiliary sequences to explore the potential interests in an open-world CDSR scenario. The denoising interest-aware network (\textbf{DIN}) not only explores explicit interests within the domain but also transfers implicit interest information across domains. With the abundant purified auxiliary sequence information, the representations of long-tailed users can be enhanced. Furthermore, through a well-designed contrastive information regularizer in the DIN and the fusion gate unit, our MACD minimizes the semantic gap and interest deviation between the target and auxiliary behaviors. }
\end{minipage}
\end{figure}

\subsection{Embedding Encoder}
\noindent\textbf{Embedding Layer.}  
To obtain the initialized sequence representations $\mathbf{S}^X=\{\mathbf{h}_{s_1}^{'X},\cdots,\mathbf{h}_{s_T}^{'X}\}$ and $\mathbf{S}^Y=\{\mathbf{h}_{s_1}^{'Y},\cdots,\mathbf{h}_{s_T}^{'Y}\}$, we utilize embedding layers $\mathrm{E}^X \in \mathbb{R}^{|\mathcal{V}^X| \times d}$ and $\mathrm{E}^Y \in \mathbb{R}^{|\mathcal{V}^Y| \times d}$, where $d$ denotes the dimension of the embeddings and $T$ is the maximum length of the interaction sequence. If the sequence length is larger than $T$, we only consider the most recent $T$ actions. If the sequence length is less than $T$, we repeatedly add a 'padding' item to the left until the length is $T$. The initialized embeddings $\mathbf{C}^X=\{\mathbf{h}_{c_1}^{'X},\cdots,\mathbf{h}_{c_{T'}}^{'X}\}$ and $\mathbf{C}^Y=\{\mathbf{h}_{c_1}^{'Y},\cdots,\mathbf{h}_{c_{T'}}^{'Y}\}$ for the auxiliary behavior sequences are obtained in the same manner. $T'$ denotes the maximum length of the auxiliary behavior sequence, which is greater than $T$. Moreover, we introduce learnable position embedding matrixes $\mathrm{P_S} \in \mathbb{R}^{T \times d}$ and $\mathrm{P_C} \in \mathbb{R}^{T' \times d}$ to improve the ordered information of sequence embeddings.

\noindent\textbf{Sequential information encoder.}  Our model-agnostic framework can directly integrate with off-the-shelf SR methods \cite{sun2019bert4rec,hidasi2015session,kang2018self}, eliminating the need to design a sequential information encoder. thus, we do not modify the sequential information encoders of SR methods further in our work.
% In this subsection, we introduce the sequential information encoders used in three baseline methods to verify the efficiency of our model. GRU4Rec \cite{hidasi2015session} uses gated recurrent units to model user action sequences. SASRec \cite{kang2018self} builds its sequential information encoder using a left-to-right Transformer language model. Similarly, the encoder of Bert4Rec \cite{sun2019bert4rec} is constructed using a bidirectional self-attention architecture. 
For simplicity, we denote their sequential information encoder as a function $\mathcal{F}=\{\mathcal{F}^X,\mathcal{F}^Y,\mathcal{F}_c^X,\mathcal{F}_c^Y\}$. Formally, the procedure is as: $\mathbf{h}_{s_1}^X,\cdots,\mathbf{h}_{s_T}^X = \mathcal{F}^X(\mathbf{h}^{'X}_{s_1},\cdots,\mathbf{h}^{'X}_{s_T})$; $\mathbf{h}_{s_1}^Y,\cdots,\mathbf{h}_{s_T}^Y = \mathcal{F}^Y(\mathbf{h}^{'Y}_{s_1},\cdots,\mathbf{h}^{'Y}_{s_T})$; $\mathbf{h}_{c_1}^X,\cdots,\mathbf{h}_{c_{T'}}^X = \mathcal{F}_c^X(\mathbf{h}^{'X}_{c_1},\cdots,\mathbf{h}^{'X}_{c_{T'}})$ and $\mathbf{h}_{c_1}^Y,\cdots,\mathbf{h}_{c_{T'}}^Y =$ 

\noindent$\mathcal{F}_c^Y(\mathbf{h}^{'Y}_{c_1},\cdots,\mathbf{h}^{'Y}_{c_{T'}})$.
% \vspace{-5pt}
% \begin{small}
% \begin{align*} 
% \mathbf{h}_{s_1}^X,\cdots,\mathbf{h}_{s_T}^X &= \mathcal{F}^X(\mathbf{h}^{'X}_{s_1},\cdots,\mathbf{h}^{'X}_{s_T}),\\
% %\end{equation}
% %\begin{equation}    
% \mathbf{h}_{s_1}^Y,\cdots,\mathbf{h}_{s_T}^Y &= \mathcal{F}^Y(\mathbf{h}^{'Y}_{s_1},\cdots,\mathbf{h}^{'Y}_{s_T}),\\
% %\end{equation}
% % \end{small}
% % \begin{small}
% %\begin{equation} 
% \mathbf{h}_{c_1}^X,\cdots,\mathbf{h}_{c_{T'}}^X &= \mathcal{F}_c^X(\mathbf{h}^{'X}_{c_1},\cdots,\mathbf{h}^{'X}_{c_{T'}}),\\
% %\end{equation}
% %\begin{equation}  
% \mathbf{h}_{c_1}^Y,\cdots,\mathbf{h}_{c_{T'}}^Y &= \mathcal{F}_c^Y(\mathbf{h}^{'Y}_{c_1},\cdots,\mathbf{h}^{'Y}_{c_{T'}}).
% \end{align*}
% \end{small}

\subsection{Denoising Interest-aware Network}
To fully leverage the information in the auxiliary behavior sequence, we propose a denoising interest-aware network that consists of an intra-domain denoising module (\textbf{IDDM}), a cross-domain denoising module (\textbf{CDDM}), and a contrastive information regularizer. The IDDM is designed to explicitly explore the user's interests, while the CDDM extracts latent interests and transfers cross-domain knowledge. To ensure that the explicit- and implicit-interest representations are consistent with user interests learned from the target behaviour sequences, we also introduce a novel contrastive information regularizer.

\noindent\textbf{Intra-domain Denoising Module.}  
Given a user $u$ with the sequence representations $\mathbf{S}$ and $\mathbf{C}$, we perform the intra-domain denoising procedure for each domain separately. We introduce an explicit-interest guided multi-head attention mechanism to efficiently extract useful information from the noisy auxiliary behavior sequence. Using a single attention head is insufficient since our objective is to extract multiple explicit interests from the users. Therefore, we modify the Multi-Head Attention mechanism described in \cite{vaswani2017attention} to eliminate redundant and unrelated information. The explicit-interest representation $\mathbf{S}^{*X} = \{\mathbf{h}_{s_1}^{*X},\cdots,\mathbf{h}_{s_T}^{*X}\}$ is obtained by $\mathbf{h}_{s_1}^{*X},\cdots,\mathbf{h}_{s_T}^{*X}= Concat(head_{1},\cdots,head_{h})W^{O} $.
where $head_{i}= Attention(QW_{i}^{Q},KW_{i}^{K},VW_{i}^{V})$ and
\begin{align} 
Q &= \mathbf{h}_{s_1}^X,\cdots,\mathbf{h}_{s_j}^X,\cdots,\mathbf{h}_{s_T}^X, \ j \in [1,T]
\label{Q1}\\
%\end{equation}
%\begin{equation}
K,V &= \mathbf{h}_{c_1}^X,\cdots,\mathbf{h}_{c_{j'}}^X,\cdots,\mathbf{h}_{c_{T'}}^X , \ j' \in [1,T']
\label{K,V}
\end{align}
\begin{align}
Attention(Q,K,V)& = softmax(\frac{QK^{T}}{\sqrt{d_k}})V,
\label{att_cal}
\end{align}
where $W^{O} \in \mathbb{R}^{d \times d}$, $W_{i}^{Q} \in \mathbb{R}^{d \times d_{k}}$, $W_{i}^{K} \in \mathbb{R}^{d \times d_{k}}$ and $W_{i}^{V} \in \mathbb{R}^{d \times d_{k}}$ are trainable matrices, $d_{k}=d/h$ and $h$ is the number of attention heads.
Similar intra-domain denoising processes with different matrix weights are performed on domain $Y$ to obtain the explicit-interest representation $\mathbf{S}^{*Y} = \{\mathbf{h}_{s_1}^{*Y},\cdots,\mathbf{h}_{s_T}^{*Y}\}$. 

\noindent\textbf{Cross-domain Denoising Module.}  
In this component, we conduct the cross-domain denoising operation to explore the users' implicit interest and transfer the cross-domain knowledge. Being similar to the intra-domain denoising module, we adopt an implicit-interest guided multi-head attention mechanism to purify the auxiliary behaviors representation. Specifically, to obtain user's refined implicit-interest representation in domain $X$, $\mathbf{\hat{S}}^{X} = \{\mathbf{\hat{h}}_{s_1}^{*X},\cdots,\mathbf{\hat{h}}_{s_T}^{X}\}$, we define the Query Q, Key K, and Value V as follows: 
\begin{align}
Q &= \mathbf{h}_{s_1}^X,\cdots,\mathbf{h}_{s_j}^X,\cdots,\mathbf{h}_{s_T}^X, \ j \in [1,T]
\label{Q}\\
%\end{equation}
%\begin{equation}
K,V &= \mathbf{h}_{c_1}^Y,\cdots,\mathbf{h}_{c_{j'}}^Y,\cdots,\mathbf{h}_{c_{T'}}^Y, \ j' \in [1,T']
\label{K,V}
\end{align}

The rest steps are the same as Eq. \ref{Q1}-\ref{att_cal}. 
The cross-domain denoising module in the domain $Y$ is similar, and we can obtain the refined implicit-interest representation in domain $Y$, $\mathbf{\hat{S}}^{Y} = \{\mathbf{\hat{h}}_{s_1}^{*Y},\cdots,\mathbf{\hat{h}}_{s_T}^{Y}\}$ correspondingly. Our motivation behind developing the denoising interest-aware network is to extract purified information as comprehensively as possible from the noisy but abundant auxiliary behavior data. To accomplish this, we consider the longer sequence of auxiliary behavior $\mathbf{C}$ as the key and value, rather than $\mathbf{S}$.

\noindent\textbf{Contrastive Information Regularizer.}   
In our intra- and cross-domain denoising modules, our aim is to extract different representations that incorporate explicit and implicit interests respectively. Moreover, we aim to minimize the variance between various interest representations of the same user and maximize the difference between that of different users by the proposed contrastive information regularizer. One of the crucial aspects of Contrastive Learning is to extract valid positive-negative pair samples from different views. In our design, we consider the composite purified interest representation of the same user $u_i$ as positive pairs, while interest representations from different users $u_i, u_j$ are considered as negative pairs. Therefore, taking domain $X$ as an example, we obtain the positive and negative sample pairs $\{\mathbf{S}^{*X}_{u_i},\mathbf{\hat{S}}^{X}_{u_i}\}$ and$\{\mathbf{S}^{*X}_{u_i},\mathbf{\hat{S}}^{X}_{u_j}\}$, where $u_i,u_j \in $ $\mathcal{U}^X$ and $i \neq j$. Formally, we define a contrastive information regularizer based on InfoNCE \cite{oord2018representation} loss, as follows:

\begin{equation}
\mathcal{L}_{cl}^{X} = \sum\limits_{u_i \in \mathcal{U}^X}-\log\frac{\exp(s(\phi(\mathbf{S}^{*X}_{u_i}),\phi(\mathbf{\hat{S}}^{X}_{u_i}))/\tau)}{\sum\limits_{u_j \in \mathcal{U}^X,i\neq j}\exp(s(\phi(\mathbf{S}^{*X}_{u_i}),\phi(\mathbf{\hat{S}}^{X}_{u_j}))/\tau)}
\end{equation}

The hyper-parameter $\tau$ regulates the smoothness of the softmax curve, while the pair-wise distance function $s(\cdot)$ evaluates the similarity between positive and negative pairs. The function $\phi(\cdot)$ applies a simple $\mathsf{Mean}$ operation to average the embedding along the temporal dimension. $\mathcal{L}_{cl}^{Y}$ is defined in a similar way. The proposed contrastive information regularizer enables our framework to learn more robust user interest-aware representations that exhibit high consistency while being capable of distinguishing personal preferences among different users.

\subsection{Fusion Gate Unit}
To represent user’s holistic preferences in each domain with distilled embeddings $\mathbf{S}, \mathbf{S}^{*}, \mathbf{\hat{S}}$, we introduce a gate unit to fuse them. Firstly, we utilize $\phi(\cdot)$ to aggregate their representation along the temporal dimension and obtain the corresponding embeddings $\mathbf{O} \in \mathbb{R}^{1 \times d}, \mathbf{O}^{*} \in \mathbb{R}^{1 \times d}, \mathbf{\hat{O}} \in \mathbb{R}^{1 \times d}$. It is a straightforward way to apply element-wise add or concatenating operation to fuse feature. However, the explicit- and implicit- interest representation $\mathbf{O}^{*} \in \mathbb{R}^{1 \times d}, \mathbf{\hat{O}} \in \mathbb{R}^{1 \times d}$ are learned from the auxiliary behavior sequences, which are less important than the sequences $\mathbf{O} \in \mathbb{R}^{1 \times d}$.
Considering different weights to fuse them can explore the full potential of embeddings. A learnable weight matrix is first learned and then conduct a gate control to fuse the embeddings, \emph{i.e.},
\begin{align}
\mathbf{H}_1^{X} &= \mathrm{Sig}(\mathbf{O}^X\mathbf{W}_1+ b_1+ \mathbf{O}^{*X}\mathbf{W}_{2}+b_2), 
\label{h1}\\
%\end{equation}
%\begin{equation}
\mathbf{O}_m^X &= \mathrm{tanh}{((1-\mathbf{H}_1^{X}) \odot \mathbf{O}^X+ \mathbf{H}_1^{X} \odot \mathbf{O}^{*X})},
\label{omx}
\end{align}
where Sig$(\cdot)$ is sigmoid function and tanh($\cdot$) is the hyperbolic tangent function. $\odot$ means the element-wise multiplication operation weighting the embeddings in a fine-grained way. $\mathbf{W}_1$, $\mathbf{W}_2 \in \mathbb{R}^{d \times d}$ and $b_{1}, b_{2} \in \mathbb{R}^{d}$ are the trainable weights and bias updated in backpropagation. Then we fuse the intermediate representation $\mathbf{O}_m^X$ with $\mathbf{\hat{O}}^X$ in a similar way as follows. 
\begin{align}
\mathbf{H}_2^{X} &= \mathrm{Sig}(\mathbf{O}_m^X\mathbf{W}_3+ b_3+ \mathbf{\hat{O}}^{X}\mathbf{W}_{4}+b_4),
\label{h1}\\
%\end{equation}
%\begin{equation}
\mathbf{\Bar{O}}^X &= \mathrm{tanh}{((1-\mathbf{H}_2^{X}) \odot \mathbf{O}_m^X+ \mathbf{H}_2^{X} \odot \mathbf{\hat{O}}^{X})},
\label{omx}
\end{align}
where $\mathbf{W}_3$, $\mathbf{W}_4 \in \mathbb{R}^{d \times d}$ and $b_{3}, b_{4} \in \mathbb{R}^{d}$ are the learnable weights and bias. Similarly, $\mathbf{\Bar{O}}^Y$ can be obtained. To generate final prediction results, building upon prior research \cite{he2017neural}, we utilize the concatenated representations of the user and item ($\mathbf{\Bar{O}}_{u_i}^X$,$\bm{v}_{j}^X$) as input to a multi-layer perceptron to generate predictions of user's preference for a target item. Formally, $\hat{y}_{u_{i}v_{j}}^{X}=\sigma( \mathrm{MLPs} (\mathbf{\Bar{O}}_{u_i}^X||\bm{v}_{j}^X))$. 
% \begin{small}
% \begin{equation}
% % \begin{array}{cc}
% \hat{y}_{u_{i}v_{j}}^{X}=\sigma( \mathrm{MLPs} (\mathbf{\Bar{O}}_{u_i}^X||\bm{v}_{j}^X))
% \label{predict1}
% % \end{array}
% \end{equation}
% \end{small}
where MLPs are the stacked fully connected layers and $\bm{v}_{j}^X$ is the target item embedding. $\sigma$ denotes the sigmoid function. The prediction layer in the domain $Y$ is similar. 

\subsection{Model Training}

%adjusted part!
Our overall training loss combines the contrastive information regularizers $L_{cl}^X$ and $L_{cl}^Y$ mentioned in $Eq.(7)$ with the classification losses $L_{cls}^X$ and $L_{cls}^Y$, using a harmonic factor $\lambda$:
\begin{equation}
\mathcal{L}= \lambda(\mathcal{L}_{cl}^{X}+\mathcal{L}_{cl}^{Y})+(1-\lambda)(\mathcal{L}_{cls}^{X}+\mathcal{L}_{cls}^{Y})
\label{cls_loss}
\end{equation}
where $L_{cls}^X$ and $L_{cls}^Y$ represent the classification loss for the recommendation task in domains $X$ and $Y$, respectively. To be specific, $L_{cls}^X$ is obtained by summing over the losses $\ell(\hat{y}_{u_{i}v_{j}}^{X},y_{u_{i}v_{j}}^{X})$ for all user-item pairs $(u_{i}, v_{j})$ in domain $X$:
\begin{equation}
\mathcal{L}_{cls}^{X}= \sum_{\substack{u_{i}\in \mathcal{U}^X,v_{j}\in \mathcal{V}^X}}	
  \ell(\hat{y}_{u_{i}v_{j}}^{X},y_{u_{i}v_{j}}^{X})
\label{cls_loss}
\end{equation}
where $\hat{y}_{u_{i}v_{j}}^{X}$ represents the predicted results and $y_{u_{i}v_{j}}^{X}$ is the corresponding ground-truth label. The loss estimator $\ell(\cdot)$ in $Eq.(14)$ is used to measure the dissimilarity between the predicted results and the ground truth labels, and it is defined as the binary cross-entropy loss shown:$\ell(\hat{y},y) = - [ y\log \hat{y} + (1-y)\log(1-\hat{y})]$. $\hat{y}$ represents the predicted results, and $y$ represents the ground truth label. Similarly, $L_{cls}^Y$ can be obtained.

\subsection{Inductive Representation Generator}
In the inference stage, we propose a novel inductive representation generator (IRG) to enhance the model's performance for cold-start users\footnote{For instance, in the scenario where domain $X$ is the target domain for item recommendation, while cold-start users only have observed items in domain $Y$.}. The core idea behind this generator is to retrieve similar user embeddings from another domain whose interests closely align with cold-start users. Specifically, for a cold-start user $u^*$ in domain $X$, we first calculate the user-user similarity based on the feature $\mathbf{\Bar{O}}^Y$. We then identify the nearest user $u_l$ and utilize their embedding $\mathbf{\Bar{O}}_{u_l}^X$ in domain $X$ to replace the embedding of the cold-start user $u^*$. An example of PyTorch-style implementation is described in Algorithm \ref{algo-irg}.

\definecolor{commentcolor}{RGB}{110,154,155}   % define comment color
\newcommand{\PyComment}[1]{\ttfamily\textcolor{commentcolor}{\# #1}}  % add a "#" before the input text "#1"
\newcommand{\PyCode}[1]{\ttfamily\textcolor{black}{#1}} % \ttfamily is the code font
\begin{algorithm*}[tb!]
\SetAlgoLined
    % \PyComment{pseudocode for our proposed inductive representation generator} \\
    \PyComment{$X$,$Y$:two different domains, N:batch size, d:embedding dimension.} \\
\PyComment{$\mathbf{\Bar{O}}^X$,$\mathbf{\Bar{O}}^Y$: user's representation after FGU, shape:[N,d]} \\
    \PyComment{ColdStartLabel$^X$,ColdStartLabel$^Y$:identify the cold-start users, shape:[N]} \\
    % \PyCode{norm\_y = Norm() * torch.norm(f\_y, dim=1, keepdim=True).t() } \PyComment{Normalization coefficient.} \\

    \PyComment{Compute Normalization Coefficient.} \\
    \PyCode{Coefficient$^Y$ $\leftarrow$ Norm($\mathbf{\Bar{O}}^Y$, Axis=1) * Transpose(Norm($\mathbf{\Bar{O}}^Y$, Axis=1)) } 
 \\
    \PyComment{Find the similar users.} \\
    \PyCode{Similarity$^Y$ $\leftarrow$  Dot\_Product($\mathbf{\Bar{O}}^Y$,Transpose($\mathbf{\Bar{O}}^Y$))/Coefficient$^Y$}  \\
    \PyComment{The index of the most similar users.} \\
    \PyCode{NearestIndex$^X$ $\leftarrow$  Argmax(Similarity$^Y$,Axis=1)}  \\
    \PyComment{Generate the cold-start users' representation from similar users.} \\
    \PyCode{ColdStartRepr$^X$  $\leftarrow$ $\mathbf{\Bar{O}}^X$[NearestIndex$^X$]}  \\
    \PyComment{Select the user representation.} \\
    \PyCode{LastRepr$^X$ $\leftarrow$ Where(ColdStartLabel$^X$, ColdStartRepr$^X$, $\mathbf{\Bar{O}}^X$)}  \\

    \PyComment{Repetition for domain $Y$.} \\

 %    \PyCode{Coefficient$^X$ $\leftarrow$ Norm($\mathbf{\Bar{O}}^X$, Axis=1) * Transpose(Norm($\mathbf{\Bar{O}}^X$, Axis=1)) } 
 % \\
 %    \PyCode{Similarity$^X$ $\leftarrow$  Dot\_Product($\mathbf{\Bar{O}}^X$,Transpose($\mathbf{\Bar{O}}^X$))/Coefficient$^X$}  \\
 %    \PyCode{NearestIndex$^Y$ $\leftarrow$  Argmax(Similarity$^X$,Axis=1)}  \\
 %    \PyCode{ColdStartRepr$^Y$  $\leftarrow$ $\mathbf{\Bar{O}}^Y$[NearestIndex$^Y$]}  \\
 %    \PyCode{LastRepr$^Y$ $\leftarrow$ Where(ColdStartLabel$^Y$, ColdStartRepr$^Y$, $\mathbf{\Bar{O}}^Y$)}  
 %    \PyComment{Subsequent prediction.} \\
\caption{Pseudocode of IRG}
\label{algo-irg}
\end{algorithm*}

\section{Experiments}
In this section, we present extensive experiments to demonstrate the effectiveness of MACD, aiming to answer the following three research questions(RQs).
\begin{itemize}
\item \textbf{RQ1}: How does MACD compare to state-of-the-art methods in open-world CDSR scenarios, particularly for long-tailed and cold-start users?

\item \textbf{RQ2}: How do the different modules of MACD contribute to the performance improvement of our method?

\item \textbf{RQ3}: Whether MACD can achieve a significant improvement when deployed on a real-world industry platform?

\item \textbf{RQ4}: When encountering open-world scenarios with varying user-item interaction density and different numbers of cold-start users, can MACD consistently achieve remarkable performance?

\item \textbf{RQ5}: How do different hyperparameter settings affect the performance of our method?
% \item \textbf{RQ5}: How do different hyperparameter settings affect the performance of our method?
\end{itemize}

Owing to space constraints, additional results and analyses—encompassing baseline descriptions, experiments on the "Phone-Elec" dataset.

% \vspace{-5pt}
\subsection{Datasets}
% \vspace{-5pt}
We conducted offline experiments on a publicly available Amazon dataset comprised of 24 distinct item domains. To generate CDSR scenarios for our experiments, we selected two pairs of domains, namely "Cloth-Sport" and "Phone-Elec". We extracted users who had interactions in both domains and then filtered out items with fewer than 10 interactions. 
We used the views behaviors as auxiliary behaviors, and both target behavior sequences and auxiliary behavior sequences were collected in chronological order. To prevent the information leak problem in previous works \cite{ma2019pi,ma2022mixed}, we then divided users into three sets: 80\% for training, 10\% for validation, and 10\% for testing in each domain. To simulate multiple open-world recommendation scenarios, we retained non-overlapping users and varied the overlapping ratio to control the number of overlapping users. In addition, we randomly selected approximately 20\% of overlapping users as cold-start users for validation and testing. 
% It should be noticed that in order to create a cold-start user, we randomly remove the sequence information in one domain of one selected overlapping user while retaining the last user-item interaction as the ground truth.
The detailed statistics of our corrected datasets in CDSR scenarios are summarized in Table \ref{data-anylis}. 

\begin{table}[tb!]
\scriptsize
% \footnotesize
%\small
\setlength{\abovecaptionskip}{0pt}
\setlength{\belowcaptionskip}{5pt}
\centering
\begin{minipage}[tb!]{0.2\textwidth}
\captionsetup{font={scriptsize}}
\caption{Statistics on the Amazon datasets. \#O: the number of overlapping users across domains. }
\end{minipage}
\hfill
\begin{minipage}[h!]{0.78\textwidth}
\scriptsize
\label{data-anylis}
\begin{threeparttable} 
\setlength\tabcolsep{1.2pt}{
{
\begin{tabular}{cc|cc|cccc|c}
\toprule
\multicolumn{2}{c|}{\textbf{Dataset}}     & $\left|\mathcal{U}\right|$       & $\left|\mathcal{V}\right|$       & $\left|\mathcal{E}\right|$  & \textbf{\#O}   & $\left|S \right|$  & $\left|C \right|$        &  \textbf{Density} \\ \midrule 
\multirow{2}{*}{Amazon} & Cloth & 76,162  & 51,350  & 888,768  & \multirow{2}{*}{28,771} &14.82 &124.83  & 0.023\%  \\
                        & Sport & 225,299 & 48,691 & 2,205,976    &     &12.31 &80.37                    & 0.020\%  \\ \midrule 
\multirow{2}{*}{Amazon} & Phone & 1,440,005  & 528,873 & 18,832,424   & \multirow{2}{*}{116,211} &15.17 &175.17  & 0.002\%  \\
                        & Elec  & 194,908 & 49,295 & 2,096,841    &   &12.31 &80.37                       & 0.022\% \\ 
\bottomrule
\end{tabular}
}}
% \begin{tablenotes}    
%         \scriptsize            
%         \item \#O: the number of overlapping users across domains. 
%       \end{tablenotes}
\end{threeparttable}
\end{minipage}
\end{table}

% \vspace{-5pt}

\subsection{Experiment Setting}
\noindent\textbf{Evaluation Protocol.}  
To test the effectiveness of our approach under an open-world environment, we varied the overlapping ratio $\mathcal{K}_{u}$ of each dataset in $\{25\%$, $75\%\}$, which corresponded to different numbers of overlapping users shared across the domains. For example, in the Amazon "Cloth-Sport" dataset with $\mathcal{K}_{u}=25\%$, we determined the number of overlapping users by applying the formula $28,771 * 0.25 = 7192$. To generate an unbiased evaluation for fair comparison \cite{krichene2020sampled}, we randomly sampled 999 negative items, which were items not interacted with by the user, along with 1 positive item that served as the ground-truth interaction. We then employed these items to form the recommended candidates for conducting the ranking test.
We used several $top$-$N$ metrics to assess the effectiveness of our approach, including the normalized discounted cumulative gain (NDCG@10) and hit rate (HR@10). Higher values of all metrics indicate improved model performance. All the experiments were conducted five times and the average values are reported.

\noindent\textbf{Compared Methods.}   
To verify the effectiveness of our model in an open-world environment, we compare
MACD with three branches of baselines including SDSR methods(BERT4Rec \cite{sun2019bert4rec}, GRU4Rec \cite{hidasi2015session}, SASRec \cite{kang2018self}), Denoising SR methods (HSD \cite{zhang2022hierarchical}, DPT \cite{zhang2023denoising}) and CDSR methods (Pi-Net \cite{ma2019pi}, DASL \cite{li2021dual}, C$^{2}$DSR \cite{cao2022contrastive}). In this paper, we do not choose conventional CDR methods as baselines because they overlook sequential information importance.

\begin{table*}[tb!]
\begin{threeparttable}
\scriptsize
%\centering
\captionsetup{font={scriptsize}}
\caption{Experimental results (\%) on the bi-directional Cloth-Sport CDSR scenario with different $\mathcal{K}_{u}$. The best results for each column are highlighted in boldface, while the second-best results are underlined.}
\label{results_cs_all}
\setlength\tabcolsep{0.1pt}{
{
\begin{tabular}{lccccccccccccccccccccccccc}
\toprule
\multirow{4}{*}{\bf Methods} & \multicolumn{4}{c}{\textbf{Cloth-domain recommendation }} & \multicolumn{4}{c}{\textbf{Sport-domain recommendation }} \\
\cmidrule(r){2-5}\cmidrule(r){6-9}
& \multicolumn{2}{c}{\ $\mathcal{K}_{u}$=25\%} & \multicolumn{2}{c}{\ $\mathcal{K}_{u}$=75\%}  & \multicolumn{2}{c}{\ $\mathcal{K}_{u}$=25\%}  & \multicolumn{2}{c}{\ $\mathcal{K}_{u}$=75\%}\\ \cmidrule(r){2-3}\cmidrule(r){4-5}\cmidrule(r){6-7}\cmidrule(r){8-9}
&NDCG@10    &HR@10    
&NDCG@10    &HR@10   &NDCG@10    &HR@10   
&NDCG@10    &HR@10    \\
% &\multirow{2}{*}{MRR} &\multicolumn{2}{c}{NDCG} & \multicolumn{3}{c}{HR} & \multirow{2}{*}{MRR} &\multicolumn{2}{c}{NDCG} & \multicolumn{3}{c}{HR}\\
% \cmidrule(r){3-4}\cmidrule(r){5-7}\cmidrule(r){9-10}\cmidrule{11-13} &  & @5 &     & @1  & @5  &     &  & @5  &     & @1  & @5  &      \\
\midrule

BERT4Rec \cite{sun2019bert4rec}
& 4.49
& 9.38
& 5.04
& 10.22
& 7.87
& 15.21
& 8.07
& 15.37 \\

BERT4Rec$\dag$ \cite{sun2019bert4rec}
& 5.41
& 11.27
& 6.53
& 13.33
& 9.05
& 17.59
& 9.90
& 18.79\\

GRU4Rec \cite{hidasi2015session}
& 5.81
& 12.03
& 6.56
& 13.22
& 9.98
& 19.20
& 10.56
& 20.07 \\

GRU4Rec$\dag$ \cite{hidasi2015session}
& 5.79
& 12.24
& 6.52
& 13.70
& 10.67
& 20.05
& 10.68
& 20.27 \\

SASRec \cite{kang2018self}
& 5.88
& 12.21
& 6.43
& 13.09
& 10.06
& 19.36
& 10.49
& 20.20 \\

SASRec$\dag$ \cite{kang2018self}
& 5.79
& 11.87
& 6.59
& 13.23
& 10.22
& 19.63
& 10.92
& 20.29 \\

\midrule

DPT$\dag$  \cite{zhang2023denoising}
& 5.39
& 11.77
& 6.30
& 13.02
& 10.25
& 20.07
& 10.40
& 20.35 \\

BERT4Rec \cite{sun2019bert4rec} + HSD $\dag$ \cite{zhang2022hierarchical}
& 5.53
& 11.41
& 6.50
& 13.47
& 9.13
& 17.65
& 10.19
& 18.98 \\

GRU4Rec \cite{hidasi2015session} + HSD $\dag$ \cite{zhang2022hierarchical}
& 5.99
& 12.40
& 6.67
& 13.77
& 10.80
& 20.23
& 10.85
& 20.46 \\

SASRec \cite{kang2018self} + HSD $\dag$ \cite{zhang2022hierarchical}
& 5.97
& 12.33
& 6.70
& 13.38
& 10.30
& 19.66
& 11.01
& 20.35  \\

\midrule

Pi-Net  \cite{ma2019pi}
& 6.18
& 12.24
& 6.60
& 13.10
& 10.03
& 19.10
& 10.74
& 19.96\\

Pi-Net$\dag$  \cite{ma2019pi}
& 6.19
& 12.45
& \underline{6.78}
& 13.51
& 10.02
& 19.72
& 10.51
& 20.54 \\

DASL \cite{li2021dual}
& 6.13
& \underline{12.67}
& 6.60
& 13.04
& 10.42
& 19.63
& 10.51
& 20.02 \\

DASL$\dag$ \cite{li2021dual}
& \underline{6.21}
& 12.50
& 6.52
& \underline{13.68}
& \underline{10.61}
& 20.09
& 10.37
& 19.81 \\

C$^{2}$DSR \cite{cao2022contrastive}
& 6.16
& 12.51
& 6.54
& 13.61
& 10.26
& 20.36
& 11.00
& 20.23 \\

C$^{2}$DSR$\dag$ \cite{cao2022contrastive}
& 6.10
& 12.62
& 6.51
& 13.64
& 10.40
& \underline{20.37}
& \underline{11.04}
& \underline{20.56} \\

\midrule

BERT4Rec \cite{sun2019bert4rec} + \textbf{MACD}
& 6.54
& 13.37
& 7.29
& 14.51
& 11.12
& 20.75
& 11.72
& 21.69 \\

GRU4Rec \cite{hidasi2015session} + \textbf{MACD}
& \bf{6.77}*
& \bf{13.47}*
& 7.41
& 14.58
& 11.14
& 20.90
& 11.63
& 21.81 \\

SASRec \cite{kang2018self} + \textbf{MACD}
& 6.69
& 13.16
& \bf{7.45}*
& \bf{14.61}*
& \bf{11.28}*
& \bf{21.02}*
& \bf{11.89}*
& \bf{21.82}* \\ 

Improvement(\%)
& 9.02
& 6.31
& 9.88
& 6.80
& 6.31
& 3.19
& 7.70
& 6.13 \\
\bottomrule
\end{tabular}
}}
\begin{tablenotes}    
        % \footnotesize     
        \centering
\item $\dag$ indicates whether the compared models utilize auxiliary behavior sequences. "*" denotes statistically significant improvements ($p$ $\leq$ 0.05), as determined by a paired t-test comparison with the second best result in each case. 
\end{tablenotes}
\end{threeparttable}
\end{table*}

\begin{table*}[tb!]
\begin{threeparttable}
\scriptsize
\centering
\captionsetup{font={scriptsize}}
\caption{Experimental results (\%) of the long-tailed and cold-start users are presented for the bi-directional Cloth-Sport CDSR scenario. The experiments are conducted five times, while we report the average values due to page limitations.
}
\label{results_cs_sub}
\setlength\tabcolsep{0.1pt}{
{
\begin{tabular}{lccccccccccccccccccccccccc}
\toprule
\multirow{4}{*}{\bf Methods} & \multicolumn{4}{c}{\textbf{Cloth-domain recommendation }} & \multicolumn{4}{c}{\textbf{Sport-domain recommendation }}  \\
\cmidrule(r){2-5}\cmidrule(r){6-9}
& \multicolumn{4}{c}{\ $\mathcal{K}_{u}$=25\%} & \multicolumn{4}{c}{\ $\mathcal{K}_{u}$=25\%}  \\
\cmidrule(r){2-5} \cmidrule(r){6-9}
& \multicolumn{2}{c}{\ long-tailed } & \multicolumn{2}{c}{\ cold-start }  & \multicolumn{2}{c}{\ long-tailed } & \multicolumn{2}{c}{\ cold-start } \\
\cmidrule(r){2-3}\cmidrule(r){4-5}\cmidrule(r){6-7}\cmidrule(r){8-9} 
&NDCG@10    &HR@10    
&NDCG@10    &HR@10   &NDCG@10    &HR@10   
&NDCG@10    &HR@10    \\
% &\multirow{2}{*}{MRR} &\multicolumn{2}{c}{NDCG} & \multicolumn{3}{c}{HR} & \multirow{2}{*}{MRR} &\multicolumn{2}{c}{NDCG} & \multicolumn{3}{c}{HR}\\
% \cmidrule(r){3-4}\cmidrule(r){5-7}\cmidrule(r){9-10}\cmidrule{11-13} &  & @5 &     & @1  & @5  &     &  & @5  &     & @1  & @5  &      \\
\midrule

BERT4Rec \cite{sun2019bert4rec}
& 4.15
& 8.81
& 5.13
& 9.70

& 5.65
& 11.55
& 6.34
& 13.52
 \\

BERT4Rec$\dag$ \cite{sun2019bert4rec}
& 5.36
& 11.00
& 5.17
& 10.21

& 5.72
& 11.72
& 6.39
& 13.73
 \\

GRU4Rec \cite{hidasi2015session}
& 5.38
& 11.38
& 4.80
& 10.00

& 6.48
& 12.01
& 6.76
& 14.63
\\

GRU4Rec$\dag$ \cite{hidasi2015session}
& 5.24
& 11.34
& 4.62
& 10.15

& 6.86
& 12.81
& 7.23
& 15.35
\\

SASRec \cite{kang2018self}
& 5.40
& 11.35
& 4.76
& 10.15

& 6.41
& 13.64
& 7.44
& 15.93
 \\

SASRec$\dag$ \cite{kang2018self}
& 5.36
& 11.34
& 4.97
& 10.00

& 6.63
& 13.58
& 7.37
& 15.00
\\

\midrule

DPT$\dag$  \cite{zhang2023denoising}
& 5.35
& 11.14
& 5.03
& \underline{10.23}

& 5.81
& 11.67
& 6.35
& 13.63
 \\

BERT4Rec \cite{sun2019bert4rec} + HSD $\dag$ \cite{zhang2022hierarchical}
& 5.31
& 11.10
& 5.01
& 10.19

& 5.77
& 11.65
& 6.32
& 13.70
\\

GRU4Rec \cite{hidasi2015session} + HSD $\dag$ \cite{zhang2022hierarchical}
& 5.27
& 11.39
& 4.64
& 10.10

& 6.90
& 12.84
& 7.25
& 15.33 \\

SASRec \cite{kang2018self} + HSD $\dag$ \cite{zhang2022hierarchical}
& 5.38
& 11.38
& 4.99
& 10.05

& 6.61
& 13.74
& 7.40
& 15.03\\

\midrule

Pi-Net  \cite{ma2019pi}
& 5.24
& 11.06
& 4.73
& 10.15

& 7.16
& 14.01
& 6.78
& 16.13
\\

Pi-Net$\dag$  \cite{ma2019pi}
& 5.37
& 11.57
& 4.53
& 10.07

& 7.21
& 13.93
& 7.05
& 16.16  \\

DASL \cite{li2021dual}
& 5.15
& 10.31
& 4.94
& 9.93

& \underline{7.62}
& 14.11
& 7.34
& 16.58
\\

DASL$\dag$ \cite{li2021dual}
& 5.43
& 10.77
& \underline{5.08}
& 9.95

& 7.43
& 14.39
& 7.51
& 16.81 \\

C$^{2}$DSR \cite{cao2022contrastive}
& 5.52
& 11.55
& 4.30
& 9.20

& 7.50
& 14.32
& 7.37
& 17.21\\

C$^{2}$DSR$\dag$ \cite{cao2022contrastive}
& \underline{5.58}
& \underline{11.56}
& 4.37
& 9.78

& 7.40
& \underline{14.45}
& \underline{7.69}
& \underline{17.29}
 \\

\midrule

BERT4Rec \cite{sun2019bert4rec} + \textbf{MACD}
& 5.90
& 11.95
& 5.56
& 10.74

& 7.52
& 14.51
& 8.09
& 17.77 \\

GRU4Rec \cite{hidasi2015session} + \textbf{MACD}
& 6.08
& \bf{12.07}*
& \bf{5.93}*
& \bf{10.91}*

& 7.78
& 14.77
& 8.13
& 17.61
\\

SASRec \cite{kang2018self} + \textbf{MACD}
& \bf{6.09}*
& 11.97
& 5.54
& 10.82

& \bf{8.02}*
& \bf{15.56}*
& \bf{8.26}*
& \bf{17.85}*
 \\ 

Improvement(\%)
& 9.14
& 4.41
& 16.73
& 6.65

& 5.25
& 7.68
& 7.41
& 3.24
 \\
\bottomrule
\end{tabular}
}}
\end{threeparttable}
\end{table*}

\noindent\textbf{Parameter Settings.}  
We set the embedding dimension $d$ to 128 and the batch size to 2048. The training epoch is fixed at 100 to obtain optimal performance and the comparison baselines employ other hyper-parameters as reported in their official code implementation.
For MACD, we set the maximum length $T$ of the main behaviour sequence $S$ to 20 and the maximum length $T'$ of the target behaviour sequence $C$ to 100. The number of
attention heads $h$ is set to 8. The harmonic factor $\lambda$ is selected from a range of values between 0.1 and 0.9 with a step length of 0.1, and the hyper-parameter $\tau$ in $\mathcal{L}_{cl}$ is set to 1.
Function $s(\cdot)$ is implemented with the $L2$ distance.
Each approach is run five times under five different random seeds and the optimal model is selected based on the highest NDCG@10 performance on the validation set, using a grid search approach.

\subsection{Performance Comparisons (RQ1)}
\noindent {\bf{Quantitative Results.} } 
Table \ref{results_cs_all} presents the performance results for HR@10 and NDCG@10 evaluation metrics across all users. Additionally, to specifically examine the performance on long-tailed users and cold-start users, we provide a separate comparison in Table \ref{results_cs_sub}.
To ensure a fair comparison, we also provided the auxiliary behavior sequences to other state-of-the-art models. With minor adjustments to the models, we utilized their designed sequential information encoder to encode the auxiliary sequences, concatenating them with the target sequence embedding before feeding them into the prediction layer.
Regarding the overall performance, our MACD framework equipped with the SDSR baselines achieve a significant improvement on Amazon datasets compared to the second-best baselines in the simulated open-world CDSR scenario, which contains a substantial number of long-tailed and cold-start users. Additionally, we make the following insightful findings: 
% \vspace{-7pt} % adjust
\begin{itemize}
    \item In most cases, both the SDSR and CDSR baselines may experience a drop in performance when using the auxiliary behaviors directly. This is because the abundance of auxiliary behaviors can bring irrelevant information to the recommender, leading to erroneous target behavior representation learning.
    \item The denoising SR method (such as HSD) improves the backbone's performance benefitting from the noiseless auxiliary behaviors and bridging the behavioral semantic gap compared to SDSR methods. But the denoising SR methods exhibit inferior performance compared with CDSR baselines since they do not consider the cross-domain information. 
    \item Though C$^{2}$DSR achieves remarkable success in most cases, the model cannot utilize auxiliary information to enhance the representation and address the semantic gap. Thus, compared to MACD, it shows an inferior performance.
    \item In most cases, incorporating auxiliary actions can enhance the performance of models on long-tailed/cold-start users, who often have sparse behavior, and auxiliary sequences can better explore their interests.
    \item Our proposed MACD consistently achieves significant performance improvements over SDSR, denoising SR, and CDSR baselines. Compared to SDSR and denoising SR baselines, we consider cross-domain information and learn the interest deviation. Additionally, compared to SDSR and CDSR baselines, we fully explore the auxiliary behavior information to better learn the representation for long-tailed and cold-start users.
\end{itemize}

\noindent {\bf{Model Efficiency.} } All comparative models were trained and tested on the same machine, which had a single NVIDIA GeForce A10 with 22GB memory and an Intel Core i7-8700K CPU with 64G RAM. Furthermore, the number of parameters for typical SASRec+HSD, DASL, C$^2$DSR, and SASRec+MACD (our approach) is within the same order of magnitude, ranging from 0.135M to 0.196M. The training/testing efficiencies of SASRec+HSD, DASL, C$^2$DSR, and SASRec+MACD (our approach) when processing one batch of samples are 4.59$\times$ $10^{-4}$s/3.70$\times$ $10^{-4}$s, 3.48$\times$ $10^{-4}$s/2.59$\times$ $10^{-4}$s, 5.46$\times$ $10^{-4}$s/3.57$\times$ $10^{-4}$s, and 4.04$\times$ $10^{-4}$s/3.41$\times$ $10^{-4}$s, respectively. In summary, our MACD approach achieves superior performance enhancement in the open-world CDSR scenario while maintaining promising time efficiency. 

\subsection{Ablation Study (RQ2)}
We conduct the following experiments on SASRec equipped with our MACD.
To verify the contribution of each key component of MACD, we conduct an ablation study with $\mathcal{K}_{u} = 25\%$ by comparing it with several variants. Based on Table \ref{abl_components}, we make the following observations:
(i) Extracting implicit interest and explicit interest is critical to enhancing the representation for users, especially for long-tailed and cold-start users. When CDDM was removed, our model is unable to collect and transfer knowledge, which significantly hurt its performance.
(ii) Without CL and FGU, the performance also dropped significantly, as the extracted user interest information from auxiliary behaviors may generate interest deviation and impair performance.
(iii) Without IRG, the cold-start users cannot obtain a well-learned inductive representation, which impairs performance. 
As such, our model equipped with all main components achieves the best performance.

\begin{table}[tb!]
\begin{minipage}[tb!]{0.4\textwidth}
\captionsetup{font={scriptsize}}
\caption{Experimental results (\%) with different model variants. $w/o$ denotes the model without the corresponding component variant. IDDM denotes the intra-domain denoising module, while CDDM denotes the cross-domain denoising module. CL is the contrastive information regularizer and FGU is the fusion gate unit. IRG denotes the inductive representation generator.}
\label{abl_components}
\scriptsize
% \centering
\end{minipage}%
\hfill
\begin{minipage}[tb!]{0.58\textwidth}
\renewcommand{\arraystretch}{1}
\scriptsize
\setlength{\tabcolsep}{1pt}
\begin{tabular}{lccccccc}
\toprule
\multirow{2}{*}{Scenarios} & \multirow{2}{*}{Metrics} & \multicolumn{5}{c}{Model variants ($w/o$)} & \multirow{2}{*}{Ours} \\
\cmidrule(lr){3-7}
 & & IDDM & CDDM & CL & FGU & IRG & \\
\midrule
\multirow{2}{*}{Cloth} & NDCG & 6.14 & 5.52 & 6.02 & 6.27 & 6.49 & \bfseries 6.69 \\
 & HR & 12.47 & 11.88 & 12.54 & 12.69 & 12.93 & \bfseries 13.16 \\
\midrule
\multirow{2}{*}{Sport} & NDCG & 9.82 & 9.59 & 10.03 & 10.60 & 10.93 & \bfseries 11.28 \\
 & HR & 19.77 & 19.44 & 19.82 & 20.39 & 20.53 & \bfseries 21.02 \\
\bottomrule
\end{tabular}
\end{minipage}
\end{table}

% study of the sparsity 
% study of the number of the cold-start users 

\subsection{Online Evaluation (RQ3)}
% \begin{minipage}[tb!]{0.5\textwidth}
Except for these offline experiments, we conduct an online A/B test on a large-scale financial platform, which consist of multiple financial domains, such as purchasing funds, mortgage loans, and discounting bills. Specifically, we select three popular domains - "Loan," "Fund," and "Account" - from the serving platform as targets for our online testing.
For the control group, we adopt the current online solution for recommending themes to users, which is a cross-domain sequential recommendation method that utilizes noisy auxiliary behaviors directly. For the experiment group, we equip our method with a mature SDSR approach that has achieved remarkable success in the past. We evaluate the results based on three metrics: the number of users who have been exposed to the service, the number of users who have clicked inside the service, and the conversion rate of the service (denoted by \# exposure, CTR, and CVR, respectively). All of the results are reported as the lift compared to the control group and presented in Table \ref{ABtest-results}. In a fourteen-day online A/B test, our method improved the average exposure by 10.29\%, the conversion rate by 6.28\%, and the CVR by 1.45\% in the three domains.
% \vspace{-10pt}
% \end{minipage}
% \hfill
% \begin{minipage}[tb!]{0.3\textwidth}
\begin{table}[tb!]
\centering
\scriptsize
% \begin{minipage}[h!]{0.15\textwidth}
\captionsetup{font={scriptsize}}
\caption{Online A/B testing results from 9.1 to 9.14, 2023}
% \end{minipage}
% \hfill
\renewcommand{\arraystretch}{1}
% \vspace{-5pt}
\label{ABtest-results}
% \footnotesize
\scriptsize
% \begin{minipage}[h!]{0.82\textwidth}
\setlength\tabcolsep{5pt}{
\begin{tabular}{lcccc} 
\toprule
              & \textbf{\# exposure} & \textbf{CTR} & \textbf{CVR}   \\ 
\midrule
Loan Domain & +10.23\%            &  +6.31\%           & +1.54\%                           \\ 

Fund Domain    & +8.51\%            &   +5.49\%          & +1.03\%                           \\ 
Account Domain    & +12.13\%            &    +7.03\%         & +1.77\%                           \\ 
\bottomrule
\end{tabular}}
% \end{minipage}
\end{table}
% \end{minipage}
% \vspace{-30pt}

\subsection{Model Analyses (RQ4)}
% detailed on appdenix b
\noindent {\bf{Discussion of the behaviour sparsity.} } 
% 25 50 75 100 \% - compared to the second-best methods
To verify the superior performance of MACD in CDSR scenarios with varying data densities, we conducted further studies by varying the data density $D_{s}$ in $\{25\%, 50\%, 75\%, 100\%\}$. As an example, in the "Cloth-Sport" task, $D_{s}=50\%$ indicates that the data densities of the "Cloth" and "Sport" domains change from 0.023\% to 0.012\% (computed as 0.023\% * 0.5 = 0.0115\%) and from 0.020\% to 0.010\% (computed as 0.020\% * 0.5 = 0.010\%), respectively. We set $\mathcal{K}_u$ to 25\%. The experimental results of our model (SASRec+MACD) compared to the second-best baseline (C$^2$DSR)\footnote{The C$^2$DSR method utilizing auxiliary behaviors obtains the second-best results.} are presented in Table \ref{compare_ds}. As expected, the performance of all models decreases with decreasing data density, as sparser data makes representation learning and knowledge transfer more challenging. Our method consistently outperforms C$^2$DSR in all sparsity experimental settings, confirming the effectiveness of our approach in the open-world environment.
\begin{table}[t]
\scriptsize
\captionsetup{font={scriptsize}}
\begin{minipage}{0.45\textwidth}
\caption{Experimental results (\%) on different density dataset. }
\renewcommand{\arraystretch}{1}
\centering
% \vspace{5pt}
\label{compare_ds}
\setlength\tabcolsep{2.5pt}{
\begin{tabular}{lccccccccc}
\toprule
\multicolumn{1}{c}{\multirow{2}{*}{Scenarios}} & \multicolumn{1}{c}{\multirow{2}{*}{$D_{s}$}}  & \multicolumn{2}{c}{Ours}                       & \multicolumn{2}{c}{C$^2$DSR}                                                   \\ 
\cmidrule(r){3-4} \cmidrule(r){5-6}
\multicolumn{1}{c}{}                           &                          & \multicolumn{1}{c}{NDCG} & \multicolumn{1}{c}{HR} & \multicolumn{1}{c}{NDCG} & \multicolumn{1}{c}{HR}  \\ 
\midrule
\multirow{4}{*}{Cloth}                    & 25\%                       &   2.87       & 7.34                & 2.33 & 6.29             \\ 
& 50\%                       & 3.42  & 9.69   & 2.94  & 8.98          \\ 
&75\%                     & 5.25 & 11.41   & 4.38  & 10.05         \\ 
&100\%                     & 6.69 &  13.16  & 6.10 & 12.62             \\ 
\cmidrule{2-6}
\multirow{4}{*}{Sport}                    & 25\%                       &   3.56       & 8.67                 & 2.84  & 8.01             \\ 
& 50\%                           & 5.21  & 12.13  & 3.49  & 10.97          \\ 
&75\%                     & 8.34 & 17.54   & 7.57  &  16.80         \\ 
&100\%                     &  11.28 & 21.02  & 10.40 & 20.37            \\  
\bottomrule
\end{tabular}}
% \end{table}
\end{minipage}
    \hfill
\begin{minipage}{0.45\textwidth}

% \begin{table}[t!]
\scriptsize
\captionsetup{font={scriptsize}}
\caption{Impact (\%) of the number of cold-start users.}
% \vspace{5pt}
\centering
\label{compare_ds}
\setlength\tabcolsep{2.5pt}{
\begin{tabular}{lccccccccc}

\toprule
\multicolumn{1}{c}{\multirow{2}{*}{Scenarios}} & \multicolumn{1}{c}{\multirow{2}{*}{$K_{cs}$}}  & \multicolumn{2}{c}{Ours}                       & \multicolumn{2}{c}{C$^2$DSR}                                                   \\ 
\cmidrule(r){3-4} \cmidrule(r){5-6}
\multicolumn{1}{c}{}                           &                          & \multicolumn{1}{c}{NDCG} & \multicolumn{1}{c}{HR} & \multicolumn{1}{c}{NDCG} & \multicolumn{1}{c}{HR}  \\ 
\midrule
\multirow{4}{*}{Cloth}                    & 5\%                       &   7.37       & 13.54                & 6.81 & 12.99             \\ 
& 20\%                       & 6.69 &  13.16  & 6.10 & 12.62                \\ 
&35\%                     & 6.25 & 12.81   & 5.88  & 12.35         \\ 
&50\%                     & 5.87 & 12.26  & 5.20 & 11.62             \\ 
\cmidrule{2-6}
\multirow{4}{*}{Sport}                    & 5\%                       &   11.56       & 21.67                 & 10.84  & 21.01             \\ 
& 20\%                           & 11.28 & 21.02  & 10.40 & 20.37           \\ 
&35\%                     & 10.74 & 20.54   & 10.17  &  19.80         \\ 
&50\%                     & 10.49 &  20.22  & 9.74 & 19.41            \\  
\bottomrule
\end{tabular}}
    \end{minipage}

\end{table}

\noindent {\bf{Discussion of the number of the cold-start users.} }
% 5 20 35 50 \% - compared to the second-best methods 
We also conducted experiments to investigate the effectiveness of our model in open-world environments with varying numbers of cold-start users. As mentioned in Section 3.1, we randomly selected partial overlapping users as cold-start users and varied the cold-start user ratio $\mathcal{K}_{cs}$ among \{5\%,20\%,35\%,50\%\}. Due to the limited page, we show the results with the metrics NDCG@10 and HR@10. From the figures, we made the following observations: (1) With the rise in the ratio of cold-start users, the performance of all models' recommendations declined, highlighting the difficulty but significance of learning embeddings for cold-start users. (2) Our model demonstrated more robust performance in making recommendations for cold-start users than the strongest baseline, C$^2$DSR. This is because our MACD generates inductive embeddings for the cold-start users and the auxiliary information is effectively utilized by our model.
% \begin{figure}[tb!]
% \centering
% \subfigure[Cloth Domain]{
% \begin{minipage}[t]{0.45\linewidth}
% \centering
% \includegraphics[width=0.99\linewidth]{Figure/abl_cs_Music_NDCG.pdf}
% %\caption{fig1}
% \end{minipage}%
% }%
% \subfigure[Cloth Domain]{
% \begin{minipage}[t]{0.45\linewidth}
% \centering
% \includegraphics[width=0.99\linewidth]{Figure/abl_cs_Music_HR.pdf}
% %\caption{fig2}
% \end{minipage}%
% } \\
% \subfigure[Sport Domain]{
% \begin{minipage}[t]{0.45\linewidth}
% \centering
% \includegraphics[width=0.99\linewidth]{Figure/abl_cs_Movie_NDCG.pdf}
% %\caption{fig2}
% \end{minipage}%
% }%
% \subfigure[Sport Domain]{
% \begin{minipage}[t]{0.45\linewidth}
% \centering
% \includegraphics[width=0.99\linewidth]{Figure/abl_cs_Movie_HR.pdf}
% %\caption{fig2}
% \end{minipage}%
% }%
% \centering
% % \vspace{-5pt}
% % \captionsetup{font={footnotesize}}
% \caption{Impact of the number of cold-start users.}
% \label{Mot2_tsne}
% \end{figure}

% S 10 15 20 25 30  cloth-sport /phone-elec 柱状图
% lambda 0.1 -0.9

\subsection{Parameter Sensitivity (RQ5)} 
This section investigates the parameter sensitivity of the sequence length $T$ and the harmonic factor $\lambda$.

\begin{figure*}[tb!]
\centering
\subfigure[Cloth Domain]{
\begin{minipage}[t]{0.25\linewidth}
\centering
\includegraphics[width=0.95\linewidth]{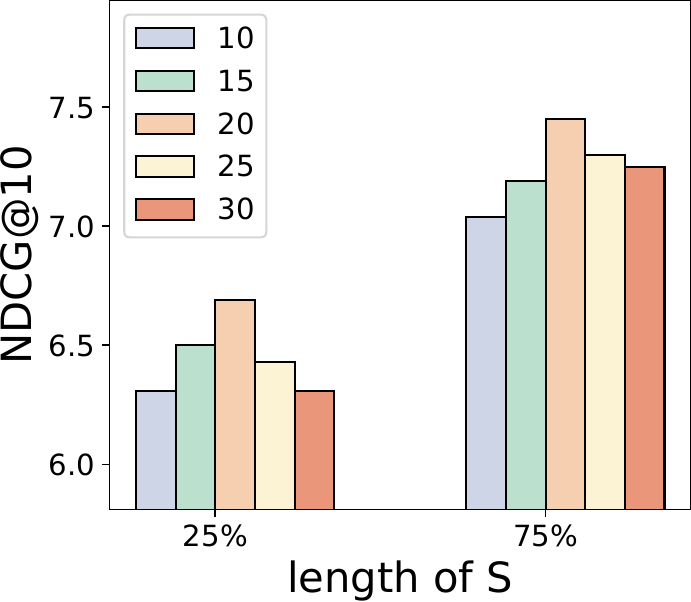}
%\caption{fig1}
\end{minipage}%
}%
\subfigure[Cloth Domain]{
\begin{minipage}[t]{0.25\linewidth}
\centering
\includegraphics[width=0.95\linewidth]{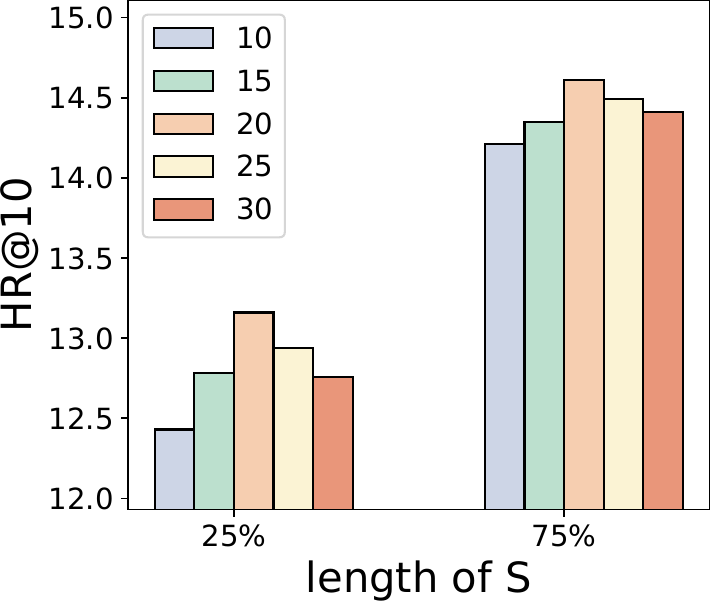}
%\caption{fig2}
\end{minipage}%
}%
\subfigure[Sport Domain]{
\begin{minipage}[t]{0.25\linewidth}
\centering
\includegraphics[width=0.95\linewidth]{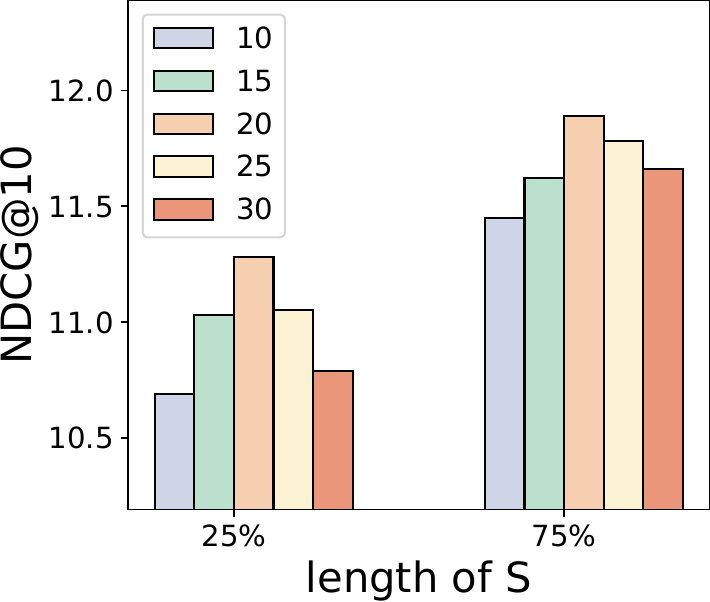}
%\caption{fig2}
\end{minipage}%
}%
\subfigure[Sport Domain]{
\begin{minipage}[t]{0.25\linewidth}
\centering
\includegraphics[width=0.95\linewidth]{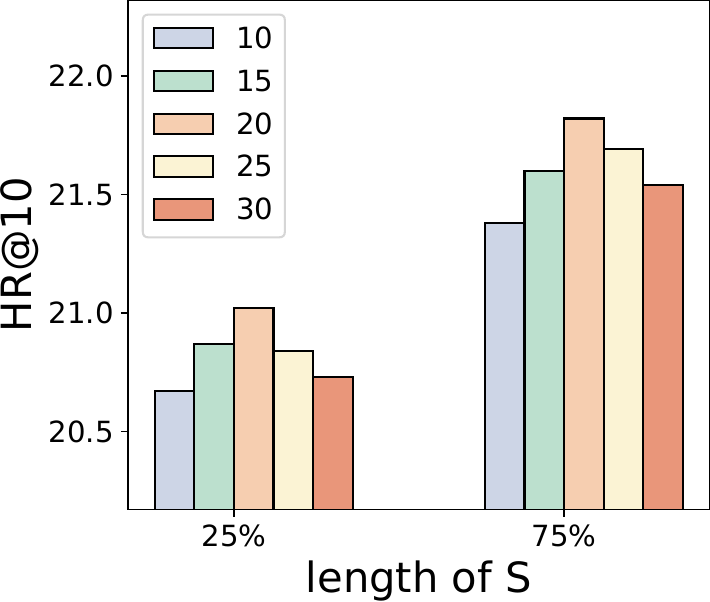}
%\caption{fig2}
\end{minipage}%
}%
\centering
% \vspace{-10pt}
% \captionsetup{font={small}}
\caption{(a)-(d) show the effect of length of sequence on model performance.}
% \vspace{-10pt}
\label{hyper_T}
\end{figure*}

For sequence length $T$, we show its "cloth" domain and "sport" domain results with overlapping ratio 25\% and 75\% in Figure \ref{hyper_T}. 
After training our model with different settings $T$ = \{10, 15, 20, 25, 30\}, one can see that our model achieves the best performance in terms of NDCG@10 and HR@10 when $T$ = 20.
When increasing $T$ from 10 to 20, the performance is gained on account of richer historical interest information. If $T$ is larger than 20, the performance will decrease.
The reason might be that padding item causes the model ignoring important information from the true user-item interaction. Therefore, we choose $T$ = 20 to better capture the user-item interaction information.

\begin{figure*}[tb!]
\centering
\subfigure[Cloth Domain]{
\begin{minipage}[t]{0.25\linewidth}
\centering
\includegraphics[width=0.95\linewidth]{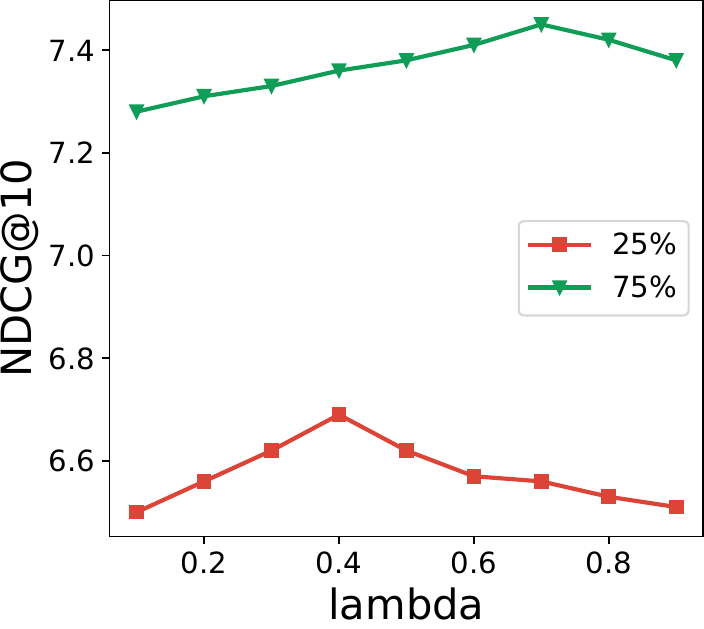}
%\caption{fig1}
\end{minipage}%
}%
\subfigure[Cloth Domain]{
\begin{minipage}[t]{0.25\linewidth}
\centering
\includegraphics[width=0.95\linewidth]{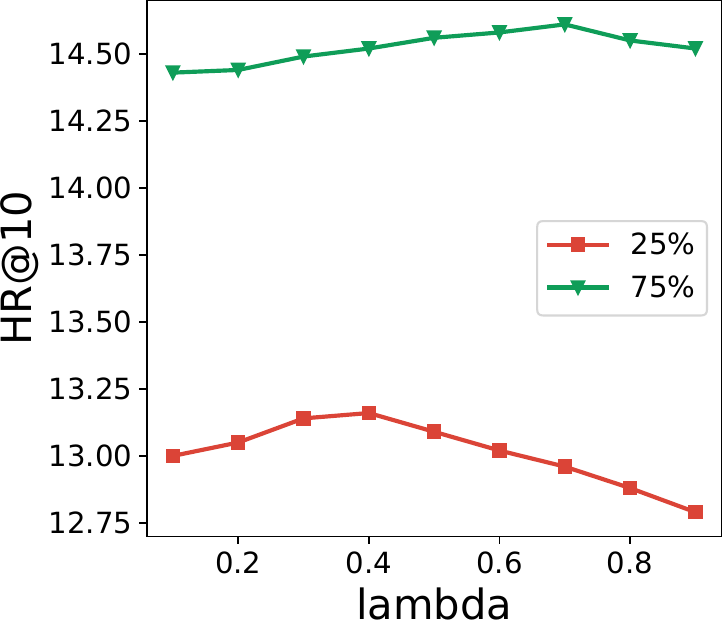}
%\caption{fig2}
\end{minipage}%
}%
\subfigure[Sport Domain]{
\begin{minipage}[t]{0.25\linewidth}
\centering
\includegraphics[width=0.95\linewidth]{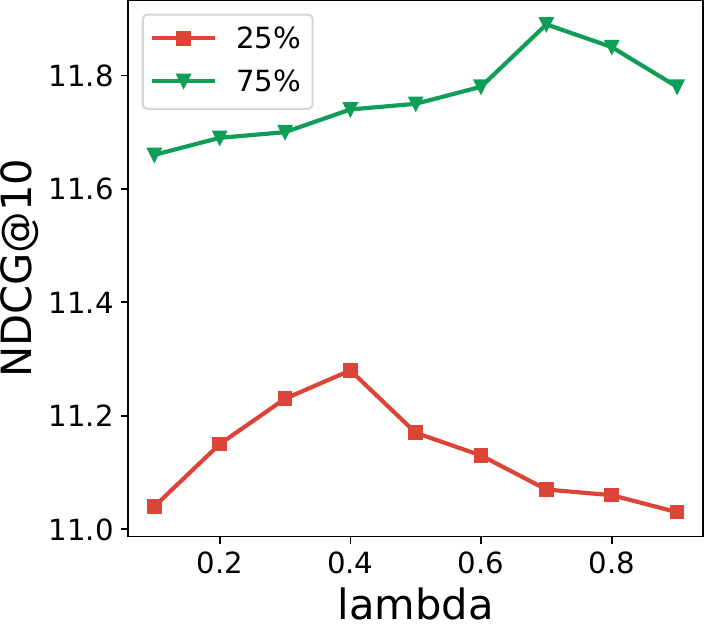}
%\caption{fig2}
\end{minipage}%
}%
\subfigure[Sport Domain]{
\begin{minipage}[t]{0.25\linewidth}
\centering
\includegraphics[width=0.95\linewidth]{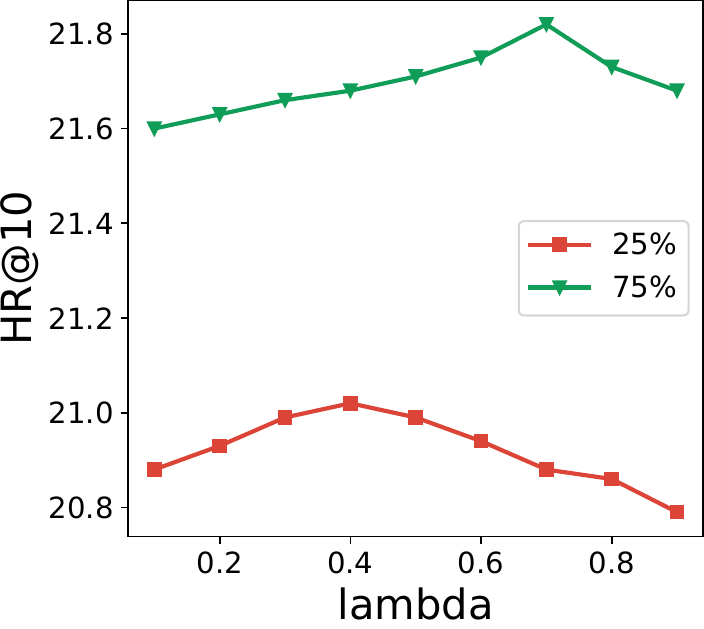}
%\caption{fig2}
\end{minipage}%
}%
\centering
% \vspace{-10pt}
% \captionsetup{font={small}}
\caption{(a)-(d) show the effect of harmonic factor $\lambda$ on model performance..}
% \vspace{-10pt}
\label{hyper_lambda}
\end{figure*}

For harmonic factor $\lambda$, Figure \ref{hyper_lambda} shows its "cloth" domain and "sport" domain prediction performance with overlapping ratio 25\% and 75\% in terms of NDCG@10 and HR@10. We report the results under $\lambda$ selected between 0.1 and 0.9 with a step length of 0.1. The curves shows that the accuracy will first
gradually increase with $\lambda$ raising and then slightly decrease.
We can conclude that when $\lambda$ approach 0, the contrastive information regularizer cannot produce positive effects. But when
$\lambda$ become too large, the contrastive loss will suppress the classification loss, which also reduces the recommendation accuracy. Empirically, we choose $\lambda = 0.4$ on
the Cloth \& Sport scenario with a $\mathcal{K}_u=25\%$ while $\lambda = 0.7$ on
the Cloth \& Sport scenario with a $\mathcal{K}_u=75\%$.

\section{Related Work}
\noindent {\bf{Sequential Recommendation}} 
% \subsection{Sequential Recommendation}
(SR) models \cite{hidasi2015session,kang2018self,xie2022contrastive,zhao2021variational,xu2024slmrec} user preferences based on historical behavioral sequences, enabling the modeling of users' dynamic interests compared to other conventional recommendation models. Various approaches have been proposed in the literature to address sequential recommendation problems.
For instance, GRU4Rec \cite{hidasi2015session} modifies classical RNN models to handle session-based recommendation problems. BERT4Rec \cite{sun2019bert4rec} utilizes a Bidirectional Encoder Representation from Transformers to better capture the diversity of users' historical behavior. SASRec \cite{kang2018self} employs self-attention mechanisms to balance the trade-off between capturing long-term semantics and addressing data scarcity issues. These models leverage different sequential models to more effectively capture the context of users' historical behavior.
However, it should be noted that these models face limitations in scenarios with data sparsity and cold-start problems in open-world recommendation systems.

\noindent {\bf{Cross-Domain Recommendation}} 
% \subsection{Cross-Domain Recommendation}
(CDR) \cite{zhu2020deep,zhao2019cross,xu2023neural}, which leverages behavior patterns from multiple domains to jointly characterize user interests, has shown great potential in addressing data sparsity and cold-start issues in single-domain recommendation system. Recent CDR studies have focused on transfer learning \cite{zhu2020graphical,ouyang2020minet,salah2021towards}, involving the design of a specific transfer module to learn a mapping function across domains and fuse pre-trained representations from each single domain. 
Furthermore, modeling domain-shared information has also drawn significant attention \cite{cao2022cross,liu2022exploiting}.
Although CDR approaches effectively incorporate rich information from relevant domains to improve performance on the target domain, conventional CDR methods still struggle to address the CDSR problems, which requires capturing sequential dependencies in users' interaction.

\noindent {\bf{Cross-Domain Sequential Recommendation}} 
% \subsection{Cross-Domain Sequential Recommendation}
(CDSR) \cite{ma2019pi,sun2021parallel,xu2023rethinking,ning2024imvae} aims to enhance sequential recommendation (SR) performance by leveraging user behavior sequences from multiple relevant domains. 
% Existing studies on CDSR can be broadly categorized into three types: RNN-based, GNN-based, and attentive learning-based models. 
Some early studies \cite{ma2019pi,sun2021parallel} employ RNNs to capture sequential dependencies and generate user-specific representations. The attentive learning-based model DASL \cite{li2021dual} uses dual attentive learning to transfer user preferences bidirectionally across domains. Moreover, C$^{2}$DSR designs sequential attentive encoders combined with contrastive learning to jointly learn inter- and intra-domain relationships. However, these methods heavily rely on data from overlapping users, which represent only a small proportion of the user pool in real-world scenarios. As a result, these methods exhibit poor performance in the open-world environment, since the insufficient representation of the long-tailed and cold-start users.

\noindent {\bf{Multi-behaviour Recommendation}} 
% \subsection{Multi-behaviour Recommendation}
% Previous work in multi-behavior recommendation
methods has investigated diverse methods for learning collective knowledge from users' behaviors, including click, add to cart, and purchase \cite{jin2020multi,xia2021graph, wei2022contrastive}. 
% From an attention-based perspective, Yang et al. propose a hypergraph-based Transformer that addresses the unbalanced distribution of multiple behavior types \cite{yang2021hyper}.
% Additionally, MATN introduces a multi-behavior Transformer Network that models the inter-dependencies among different types of user behaviors.
% Taking a graph neural network approach, 
Jin et al. employ graph convolutional networks to capture behavior-aware collaborative signals \cite{jin2020multi}, while CML introduces a self-supervised learning method for multi-behavior recommendation \cite{wei2022contrastive}. 
However, these methods often neglect the dynamism of multi-behavior relations and user interests, as they primarily focus on static recommendation scenarios.
In recent studies, DPT proposes a three-stage denoising and prompt-tuning paradigm to mitigate the noise in auxiliary behavior data \cite{zhang2023denoising}. Nevertheless, existing multi-behavior sequential recommendation techniques that concentrate on a single domain are insufficient for cross-domain scenarios, as they fail to effectively capture the divergent user preferences across domains and overlook the information transfer between domains.

\section{Conclusion}
% \subsection{Conclusion}
In this work, we propose a model-agnostic contrastive denoising framework that can be integrated with most off-the-shelf SDSR methods. To enhance open-world CDSR performance by capturing comprehensive interest information, we integrate auxiliary behavior data to refine user embeddings, particularly for long-tailed and cold-start users. However, leveraging auxiliary behaviors without adjustment can lead to semantic discrepancies from target behaviors and inaccuracies in user interest learning. To mitigate this, we introduce a denoising interest-aware network with a contrastive information regularizer for precise latent interest extraction and cross-domain knowledge transfer. We also devise a parameter-free inductive representation generator to effectively identify analogous representations for cold-start users. Our model, rigorously evaluated on public datasets, demonstrates exceptional performance in open-world CDSR, a finding corroborated by an A/B test on a large-scale financial platform.


% \bibliography{IEEEabrv}

\end{document}